\pdfoutput=1

\documentclass[aps,pra,10pt,twocolumn,showpacs,superscriptaddress,nofootinbib]{revtex4}
\usepackage{amsmath}
\usepackage{latexsym}
\usepackage{amssymb}
\usepackage{graphics,epstopdf}
\usepackage[colorlinks=true, citecolor=blue, urlcolor=blue]{hyperref}

\begin{document}
\title{Measurement incompatibility and Channel Steering}

\author{Manik Banik}
\email{manik11ju@gmail.com}
\affiliation{Physics and Applied Mathematics Unit, Indian Statistical Institute, 203 B. T. Road, Kolkata 700108, India.}

\author{Subhadipa Das}
\email{sbhdpa.das@bose.res.in }
\author{A. S. Majumdar}
\email{archan@bose.res.in}
\affiliation{S. N. Bose National Centre for Basic Sciences
Block JD, Sector III, Salt Lake, Kolkata 700098, India}

\begin{abstract}
 Incompatible measurements in quantum theory always lead to 
Einstein-Podolsky-Rosen (EPR)-Schr\"{o}dinger steering. Channel steering which
is a generalized notion of EPR-Schr\"{o}dinger steering, has been introduced
recently. Here we establish a connection between lack of joint
measurability and channel steering.


\end{abstract}

\maketitle

One of the important features of quantum theory is that not all measurements 
are compatible, i.e., they cannot be carried out simultaneously.
Such a counter intuitive aspect  makes quantum physics distinct from classical
physics. This property is intimately connected to central tenets in the theory, 
such as Heisenberg's uncertainty principle \cite{Heisenberg'27}, and Bohr's 
complementarity principle \cite{Bohr'28}. In the case of von Neuman 
measurements (projective measurements), compatibility is uniquely
captured by the notion of commutativity. Non-commuting observables in quantum 
mechanics do not admit unambiguous joint measurement \cite{Varadarajan'85}. 
With the introduction of generalized measurements, i.e., positive 
operator-valued measures (POVMs) \cite{Kraus'83,Nielsen'10}, it was shown 
that observables which do not admit perfect joint measurement, may allow   
sufficiently fuzzy joint 
measurement \cite{Busch'85}. Since for general measurements 
there is no unique notion of compatibility, here we focus on the 
well-defined criterion of joint measurability \cite{Busch'96}.

The optimal degree of unsharpness that guarantees joint measurement for all 
possible pairs of dichotomic observables of a theory may be considered as the 
degree of complementarity of the theory, which quantitatively binds the amount 
of optimal violation of the Bell-Clauser-Horne-Shimony-Holt (CHSH) inequality 
for any theory which satisfies
the no-signaling principle \cite{Banik'13}. It is also known
 that any set of two incompatible POVMs with binary outcomes may lead to a
violation of the   Bell-CHSH inequality \cite{Wolf'09}. However, this may not 
be extended to the general case of an arbitrary number of POVMs with 
arbitrarily many outcomes, since pairwise joint measurability does not imply 
full joint measurability in general \cite{Kraus'83}. On the other hand, it has 
been shown recently that measurement incompatibility in quantum theory always 
leads to 
EPR-Schr\"{o}dinger steering \cite{Quintino'2014}. It has been further shown
by one of the authors of this article that the connection between measurement incompatibility
and steering holds in a class of tensor product theories rather than just Hilbert space quantum mechanics \cite{Banik'2015}.  

Steering \cite{EPR'35} refers to the scenario where one party, 
usually called Alice, wishes to convince the other party, called Bob, that she 
can steer the state at Bob’s side by making measurements on her side. 
 Steering has attracted much attention in recent years with the formulation
 of its information theoretic perspective \cite{Wiseman'07}, as well 
as the subsequent 
development \cite{walborn'11} and 
applications \cite{schn'13} of steering inequalities. Experimental
demonstrations of steering have followed using different settings and loophole
free arrangements  \cite{saunders'10}. Practical 
applications of steering have been suggested in one-sided device-independent 
quantum key distribution \cite{Branciard'12} and sub-channel 
discrimination \cite{Piani'14}. A resource theory of steering has also been 
proposed \cite{Gallego'14}.
For the present purpose it is important to note that a set of POVMs in  
finite dimensions is not jointly measurable if and only if the set can be used 
to show the steerability of some quantum state \cite{Quintino'2014}.

Recently, the notion of steerability of quantum channels has been introduced
by  Piani \cite{Piani'14(1)}, generalizing  EPR-Schr\"{o}dinger steerability. 
Consider that there is a quantum transformation (a quantum channel) from Charlie
 to Bob, which may applied/used by Bob. Such transformation is in general 
noisy with information leaking to the environment (Alice). The relevant 
question here is the following: is Alice coherently connected to the 
input-output of the channel, or can she be effectively considered just a 
``classical bystander", with at most access to classical information about 
the transformation that affected the input of the channel ? Steerability of a 
channel has been defined as the possibility for Alice to prove to 
Bob that she is not a ``classical bystander", i.e., she is 
coherently connected with 
the input-output of the channel from Charlie to Bob. The way for Alice to
prove so is  by informing Bob of the choice of measurements performed by her 
 and their outcomes.
In this work we show that Alice is required to perform incompatible 
measurements in order to demonstrate channel steering. 

We begin by first briefly discussing the mathematical framework of POVMs
required for  studying the notion of steerability for channels as introduced
by Piani \cite{Piani'14(1)} as a generalization of the EPR-Schr\"{o}dinger 
steering scenario. A POVM consists of a collection of operators $\{M_{a|x}\}_a$ 
which are positive, $M_{a|x}\ge 0~\forall~a$, and sum up to the identity, 
$\sum_aM_{a|x}=\mathbf{1}$. Here $a$ denotes measurement outcome and $x$ 
denotes  measurement choice. A POVM may be realized physically by first 
letting the physical system interact with an auxiliary system and then 
measuring an ordinary observable on the auxiliary system. Let 
$\{M_{\vec{a}}\}$ be a set of measurements with outcome $\vec{a}=[a_{x=1},a_{x=2}, . . . ,a_{x=m}]$, where $a_x\in\{0,1,..,n\}$ is the outcome of the $x^{th}$ 
measurement.
 A set of $m$ 
POVMs $\{M_{a|x}\}_a$ is called jointly measurable if 
\begin{equation}
M_{\vec{a}}\ge 0,~~\sum_{\vec{a}}M_{\vec{a}}=\mathbf{1},~~\sum_{\vec{a}\backslash a_x}M_{\vec{a}}=M_{a|x}~\forall~x,
\label{joint-meas}
\end{equation}
where $\vec{a}\backslash a_x$ stands for the elements of $\vec{a}$ except for 
$a_x$. All POVM elements $M_{a|x}$ are recovered as marginals of the 
observable $M_{\vec{a}}$.

The EPR-Schr\"{o}dinger steering experiment can be completely characterized by 
specifying an `assemblage' $\{\sigma_{a|x}\}_{a,x}$, the set of sub-normalized 
states which Alice steers Bob into, given her choice of measurement $x$ and 
outcome $a$.  She can choose to perform one measurement from a set of $m$ 
choices, each of which has $n$ possible outcomes. The assemblage encodes  
the conditional probability distribution of her outcomes given her inputs 
$p(a|x) = \mbox{Tr}(\sigma_{a|x})$, as well as the conditional states prepared 
for Bob given Alice's input and outcome $\hat{\sigma}_{a|x} = \sigma_{a|x}/p(a|x)$. All valid assemblages satisfy the consistency requirements,  $\sum_a\sigma_{a|x}=\sum_a\sigma_{a|x'},~\forall~x\ne x'$ and $\mbox{Tr}(\sum_a\sigma_{a|x})=1$. 
This encodes the fact that Alice cannot signal to Bob, and that without any 
knowledge about Alice, Bob still holds a valid quantum state. We denote this 
set of valid assemblages as $\Sigma^S$.

Assemblages which can be created via classical strategies  (without using 
entanglement) are called unsteerable and denoted as $\Sigma^{US}$. Unsteerable 
assemblages can be expressed in the form
\begin{eqnarray}
\sigma_{a|x}=\sum_{\lambda}p(a|x,\lambda)\sigma_{\lambda},~~\forall~a,x
\end{eqnarray}
such that $\mbox{Tr}(\sum_{\lambda}\sigma_{\lambda}=1),~~\sigma_{\lambda}\ge 0~\forall\lambda$,
where $\lambda$ is a (classical) random variable held by Alice, $p(a|x,\lambda)$
are conditional probability distributions for Alice, and $\sigma_{\lambda}$ are 
the states held by Bob. Collection of unsteerable assemblages form a convex set
 \cite{Pusey'13}. Any assemblage that cannot be written in the 
above form is called steerable. For such assemblages there is no classical 
explanation as to how the different conditional states held by Bob could 
 be prepared by Alice.

EPR and Schr\"{o}dinger  \cite{EPR'35} observed that by 
performing measurements on her part of entangled quantum state shared with Bob,
 Alice can remotely prepare steerable assemblages on Bob's side. Let us denote 
the measurement assemblage on Alice's side as $\{M^A_{a|x}\}_{a,x}$, where 
$M^A_{a|x}\ge 0~\forall~a,x$ and $\sum_aM^A_{a|x}=\mathbf{1}~\forall~x$. 
This measurement assemblage whenever performed on Alice's part of a bipartite 
quantum state $\rho^{AB}$ shared between Alice and Bob, gives rise the to 
the sub-states assemblage $\{\sigma_{a|x}\}_{a,x}$ with $\sigma_{a|x}=\mbox{Tr}_A(M^A_{a|x}\otimes\mathbf{1}^B\rho^{AB})$ and $\sum_a\sigma_{a|x}=\mbox{Tr}_A(\rho^{AB})$ on Bob's side. Though Schr\"{o}dinger 
pointed out steerability of bipartite 
pure entangled states in the very early days of quantum theory, it took a long 
time to establish that there exist mixed entangled states which exhibit this 
property \cite{Wiseman'07}.

A quantum channel $\Lambda^{S\rightarrow S'}:\mathcal{D}(\mathcal{H}_S)\longrightarrow\mathcal{D}(\mathcal{H}_{S'})$ is a completely-positive trace-preserving 
linear map \cite{Wilde'13}, where $S$ and $S'$, 
respectively, are the input and output quantum systems of the channel, 
and $\mathcal{H}_*$ denotes the Hilbert space associated with the system.  
$\mathcal{D}(\mathcal{H}_*)$ denotes the set of density operators acting 
on $\mathcal{H}_*$.  We will denote a channel simply by $\Lambda$, whenever it 
is not required to specify the input-output system. 
The collection of completely-positive maps $\Lambda_a$ is called an instrument 
$\mathcal{I}$, if $\sum_a\Lambda_a$ is a channel. In such a case, each 
$\Lambda_a$ is a subchannel, i.e., a completely positive trace-non-increasing 
linear map. A channel assemblage $\mathcal{CA}:= \{\mathcal{I}_x\}_x=\{\Lambda_{a|x}\}_{a,x}$ for a channel $\Lambda$ is a collection of instruments $\mathcal{I}_x$ for $\Lambda$, i.e., $\sum_a\Lambda_{a|x}=\Lambda$ for all $x$. 

Consider a noisy quantum channel from $C$ to $B$,  `leaking' information to 
the environment. Suppose that Alice has access to some part $A$ of said 
environment. The situation can be modeled by quantum broadcast channels with 
one sender and two receivers \cite{Yard'11}.
This broadcast channel $\Lambda^{C\rightarrow AB}$ is a channel extension of the 
given quantum channel
$\Lambda^{C\rightarrow B}$.
A channel extension $\Lambda^{C\rightarrow AB}$ of a channel
$\Lambda^{C\rightarrow B}$ is called an incoherent extension if there exists an 
instrument $\{\Lambda^{C\rightarrow B}_{\lambda}\}_{\lambda}$ with $\sum_{\lambda}\Lambda^{C\rightarrow B}_{\lambda}=\Lambda^{C\rightarrow B}$, and normalized (unit trace) 
quantum states $\{\sigma^A_{\lambda}\}$, such that
\begin{equation}
\Lambda^{C\rightarrow AB}=\sum_{\lambda}\Lambda^{C\rightarrow B}_{\lambda}\otimes\sigma^A_{\lambda}.
\end{equation}
A channel extension is called a coherent extension if it is not incoherent.

We now address the issue as to under
what circumstances is Alice coherently connected to the input-output of the 
channel. In such a  case  the map from $C$ to $AB$ is a coherent extension of 
the channel from $C$ to $B$.   Steerability of a channel extension is defined 
as the possibility for Alice to prove to Bob that she is not a classical 
bystander, or in other words that the leakage of information from $C$ to $A$ 
cannot be described in terms of a classical channel. 
As in the case of EPR-Schr\"{o}dinger steering, here Alice is untrusted in the
sense that we have no knowledge of either the state that Alice holds, or the 
measurements she performs.
Note that one does not need to rely on the details/implementation of Alice's 
measurements, i.e., the situation is device-independent on Alice's side. Thus, 
the verification procedure does not require Bob to trust Alice's measurement 
devices. Every choice of measurement by Alice corresponds to a different 
decomposition into subchannels of the channel used by Bob. A channel 
assemblage $\mathcal{CA}=\{\Lambda_{a|x}\}_{a,x}$  is  unsteerable if 
there exists an instrument $\{\Lambda_{\lambda}\}_{\lambda}$, and conditional 
probability distributions
$p(a|x,\lambda)$, such that
\begin{equation}
\Lambda_{a|x}=\sum_{\lambda}p(a|x,\lambda)\Lambda_{\lambda},~~~\forall~a,x.
\end{equation}  
An unsteerable channel assemblage is denoted as $\Lambda^{US}$. A channel 
assemblage is steerable if it cannot be expressed in the above form.
In the following we show that Alice is able to produce a steerable 
channel assemblage if and only if the measurements she performs are 
incompatible.

{\bf Theorem:} The channel assemblage $\{\Lambda^{C\rightarrow B}_{a|x}\}_{a,x}$ for a channel $\Lambda=\Lambda^{C\rightarrow B}$, with $\Lambda^{C\rightarrow B}_{a|x}=\mbox{Tr}_A(M^A_{a|x}\Lambda^{C\rightarrow AB}[*])$, is unsteerable for any channel extension $\Lambda^{C\rightarrow AB}$ of $\Lambda^{C\rightarrow B}$ if and only if the set 
of POVMs $\{M^A_{a|x}\}_x$ applied by Alice on $A$ is jointly measurable.

{\bf Proof}:
We first prove that joint measurability implies no channel steering.
Let, $\{M^A_{a|x}\}_{a,x}$ be jointly measurable, with the joint measurement 
operator denoted as $M^A_{\vec{a}}$, i.e.,
\begin{equation*}
M^A_{\vec{a}}\ge 0,~~\sum_{\vec{a}}M^A_{\vec{a}}=\mathbf{1},~~\sum_{\vec{a}\backslash a_x}M^A_{\vec{a}}=M^A_{a|x},
\end{equation*}
where $\vec{a}=[a_{x=1},a_{x=2},...,a_{x-m}]$.
Our aim is to show that the channel assemblage $\{\Lambda^{C\rightarrow B}_{a|x}\}_{a,x}$ resulting from the measurement assemblage $\{M^A_{a|x}\}_{a,x}$ on Alice's 
side for any channel extension (incoherent as well as coherent) 
$\Lambda^{C\rightarrow AB}$ of $\Lambda^{C\rightarrow B}$ is unsteerable, or in other
words, there exists an instrument $\{\Lambda^{C\rightarrow B}_{\lambda}\}_{\lambda}$, 
with $\sum_{\lambda}\Lambda^{C\rightarrow B}_{\lambda}=\Lambda^{C\rightarrow B}$, and a 
conditional probability distribution $p(a|x,\lambda)$, such that 
\begin{equation}
\Lambda^{C\rightarrow B}_{a|x}=\sum_{\lambda}p(a|x,\lambda)\Lambda^{C\rightarrow B}_{\lambda}, ~~~\forall~~a,x.
\end{equation} 
Let, $\lambda=\vec{a}$,
$\Lambda^{C\rightarrow B}_{\lambda}=\Lambda^{C\rightarrow B}_{\vec{a}}=\mbox{Tr}_A(M^A_{\vec{a}}\Lambda^{C\rightarrow AB}[*])$ and
$p(a|x,\lambda)=p(a|x,\vec{a})=\delta_{a,a_x}$.
Clearly we have,
\begin{eqnarray}
\sum_{\lambda}p(a|x,\lambda)\Lambda^{C\rightarrow B}_{\lambda}&=& \sum_{\vec{a}}p(a|x,\vec{a})\Lambda^{C\rightarrow B}_{\vec{a}}\nonumber\\
&=& \sum_{\vec{a}}\delta_{a,a_x}\mbox{Tr}_A(M^A_{\vec{a}}\Lambda^{C\rightarrow AB}[*])\nonumber\\
&=& \mbox{Tr}_A(\sum_{\vec{a}}\delta_{a,a_x}M^A_{\vec{a}}\Lambda^{C\rightarrow AB}[*])\nonumber\\
&=& \mbox{Tr}_A(M^A_{a|x}\Lambda^{C\rightarrow AB}[*])\nonumber\\
&=& \Lambda^{C\rightarrow B}_{a|x}=(\Lambda^{C\rightarrow B})^{US}.
\end{eqnarray}
In Ref.\cite{Piani'14(1)} it was shown that every unsteerable channel 
assemblage can be thought as arising from an incoherent channel extension. 
We can therefore conclude that by performing compatible measurements Alice 
cannot convince Bob that she is coherently connected with the input-output of 
the noisy channel applied by Bob.

We now prove the converse of the above result that
if the channel assemblage $\{\Lambda^{C\rightarrow B}_{a|x}\}_{a,x}$ for a channel $\Lambda=\Lambda^{C\rightarrow B}$ with $\Lambda^{C\rightarrow B}_{a|x}=\mbox{Tr}_A(M^A_{a|x}\Lambda^{C\rightarrow AB}[*])$ is unsteerable for any channel extension, then 
the measurement assemblage $\{M^A_{a|x}\}_{a,x}$ applied by Alice is jointly 
measurable. In order to do so we  use the Choi-Jamio\l{}kowski representation 
\cite{Jamio'72} of channels. The Choi-Jamio\l{}kowski  isomorphic 
operator of the channel $\Lambda^{C\rightarrow AB}$ is given by
\begin{equation}
J_{C'AB}(\Lambda^{C\rightarrow AB}):=\Lambda^{C\rightarrow AB}[\psi^{CC'}_+],
\end{equation}
where $\psi^{CC'}_+$ is the density matrix corresponding to a fixed maximally 
entangled state of systems $C$ and $C'$, with $C'$ a copy of $C$.

The measurement assemblage $\{M^A_{a|x}\}_{a,x}$ performed by Alice on her part 
of the extended channel $\Lambda^{C\rightarrow AB}$ results in the channel 
assemblage $\{\Lambda^{C\rightarrow B}_{a|x}\}_{a,x}$ for the channel $\Lambda^{C\rightarrow B}$, where $\Lambda^{C\rightarrow B}_{a|x}=\mbox{Tr}_A(M^A_{a|x}\Lambda^{C\rightarrow AB}[*])$ with the Choi-Jamio\l{}kowski operator $J_{C'B}(\Lambda^{C\rightarrow B}_{a|x})$. In \cite{Piani'14(1)}, it has also been proved that the channel 
extension of a channel is steerable if and only if its Choi-Jamio\l{}kowski 
operator is steerable.
Now, if the channel assemblage $\{\Lambda^{C\rightarrow B}_{a|x}\}_{a,x}$ is 
unsteerable,  there exists an instrument $\{\Lambda^{C\rightarrow B}_{\lambda}\}_{\lambda}$, with $\sum_{\lambda}\Lambda^{C\rightarrow B}_{\lambda}=\Lambda^{C\rightarrow B}$, 
and conditional probability distribution $p(a|x,\lambda)$, such that 
$\Lambda^{C\rightarrow B}_{a|x}=\sum_{\lambda}p(a|x,\lambda)\Lambda^{C\rightarrow B}_{\lambda}$ for all $a,x$. Clearly, the Choi-Jamio\l{}kowski operator assemblage 
$\{J_{C'B}(\Lambda^{C\rightarrow B}_{a|x})\}_{a,x}$ of the Choi-Jamio\l{}kowski 
operator $J_{C'B}(\Lambda^{C\rightarrow B})$ is also unsteerable, i.e.,
\begin{equation}
J_{C'B}(\Lambda^{C\rightarrow B}_{a|x})=\sum_{\lambda}p(a|x,\lambda)J_{C'B}(\Lambda^{C\rightarrow B}_{\lambda}),~~~\forall~a,x.
\end{equation} 
It now follows from the result of Refs.\cite{Quintino'2014} that
one can construct joint measurements for the measurement assemblage $\{M^A_{a|x}\}_{a,x}$. $~~~~~~~~~~~~~~~~~~~~~~~~~~~~~~~~\blacksquare$

To, summarize, in the present work we have studied the link between
lack of joint measurability and channel steering.
An important connection was  established earlier between 
EPR-Schr\"{o}dinger steering and the joint measurement of quantum observables. 
It was shown in Refs.\cite{Quintino'2014} that incompatible 
measurements are needed to be performed for demonstrating EPR-Schr\"{o}dinger 
steering. A generalization of the notion of EPR-Schr\"{o}dinger steering 
has been introduced recently through the concept of channel steering 
\cite{Piani'14(1)}.  Here one considers a noisy quantum transformation
or channel between two parties (say, Charlie and Bob), leaking some 
information to the environment which is accessible to another party (say, 
Alice). The task of channel steering is for Alice to convince Bob that she is 
coherently connected to the input-output of the channel. In this work we 
have shown that Alice needs to perform incompatible measurements to 
succeed in her aim. By performing measurements that are
jointly measurable Alice succeeds to produce only unsteerable channel 
assemblages of the noisy channel from Charlie to Bob.
Our result establishes that non-joint measurability and channel steering
imply each other. The connection between the two may have implications
 \cite{pussey'15} for a resource theory of measurement incompatibility.

{\emph Acknowledgments:}  A.S.M.  acknowledges support from the project SR/S2/LOP-08/2013 of DST, India.

\end{document}